# Spinopelvic Anatomic Parameters Prediction Model of NSLBP based on data mining

Cheng hua [1]

## Abstract

**Objective**: The purpose of this study is to perform analysis through the low back pain open data set to predict the incidence of non-specific chronic low back pain (NSLBP) to obtain a more accurate and convenient sagittal spinopelvic parameter model.

**Methods**: The logistic regression analysis and multilayer perceptron(MLP) algorithm is used to construct a NSLBP prediction model based on the parameters of the spinopelvic parameters from open data source.

**Results:** Degree of spondylolisthesis(DS), Pelvic radius (PR), Sacral slope (SS), Pelvic tilt (PT) are four predictors screened out by regression analysis that have significant predictive power for the risk of NSLBP. The overall accuracy of the equation prediction model is 85.8%.The MLP network algorithm determines that DS is the most powerful predictor of NSLBP through more precise modeling. The model has good predictive ability of 95.2% of accuracy.

**Conclusions**: MLP models play a more accurate role in the construction of predictive models. Computer science is playing a greater role in helping precision medicine clinical research.

## Introduction

Pain, muscle tension or stiffness localized below the costal margin and above the inferior gluteal folds with or without sciatica[1] are considered as the non-specific chronic low back pain (NSCLBP) which cause substantial burden to patients and society[2] and affected by a combination of physical, psychological, environmental, cultural and social factors[3]. The possible origin of pain sources of NSLBP include variable combinations of degenerative alterations in one or more discs, facet joints, and/or ligaments, with or without regional and/or global alterations in spinal alignment[4]. A lack of coordination of the muscles that support the spine is one of the proposed mechanisms for the onset and/or persistence of NSLBP[5]. Initial non-pharmacological treatment includes educations of self-management and normal activities or exercises resumption, and psychological programs for those with persistent symptoms[6].Treatment focuses on reducing pain and its consequences became the main treatment for NSLBP because of its unknow pathoanatomical cause. Analgesic medicines, non-pharmacological therapies, first-line treatments, such as rest, opioids, spinal injections and surgery, timely review also are parts of the management[7].

---

[1] **Title:** *Medical Doctor;* **Institute:** *Sports Science school of Lingnan Normal University, Zhanjiang, Guangdong, China.* **Email:** *trernghwhuare@aliyun.com; Funding: Science and Technology Bureau (Zhanjiang City) technological guidance special project(No. 2017A01014.)*

However, determining the multifactorial cause of NSLBP is complicated and anatomical abnormalities are common in the spine and may be clinical asymptoms[8]. Standing radiographs to assess sagittal spinal alignment and MRI scan to determine the mechanism of injury could be beneficial to alternative treatment options to decrease the pain and functional limitations[9]. Although specific exercise training therapies are recommended to treat persistent NSLBP[10],they were not more cost-effective compared with other interventions for low back pain[11]. As for the aspect of decreasing pain and disability in people with chronic low back pain, multidisciplinary biopsychosocial rehabilitation interventions were more effective than usual care (moderate quality evidence) and physical treatments (low quality evidence)[12].

Spinopelvic mobility concerns about the complex interaction of hip, pelvis, and spine [13]. Acetabular anteversion, pelvic tilt, and lumbar lordosis coordinated biomechanically among spinopelvic motion[14]. Larger lumbar lordosis due to larger pelvic incidence may be a risk factor for the development of standing-induced low back pain[15]. Normal spinopelvic parameters change along with the posture like from standing to sitting[16-18]. Sagittal plane deformities and global spinal alignment have in the generation of pain and disability. Restoration or maintenance of physiological sagittal spinal alignment is imperative to achieve good clinical outcomes.

The purpose of this study is to perform analysis through the low back pain open data set to predict the incidence of NSLBP to obtain a more accurate and convenient sagittal spinopelvic parameter model.

## Methods

**Data processing**

Our data source is from the open dataset of *Kaggle*. Data contains parameters of 310 observations. There are 13 attributes for analyzation purposes, which 12 are numeric predictors ($X_1, X_2, \ldots, X_{12}$) and 1 is binary class attribute (0=Abnormal, 1=Normal) with no demographics(*Tab.1*).

Tab. 1  *Data discerption*

| | | | type | Assignment |
|---|---|---|---|---|
| Predictors (X) | $X_1$ | Pelvic Incidence | numeric, float64 | (°) |
| | $X_2$ | Pelvic Tilt | numeric, float64 | (°) |
| | $X_3$ | Lumbar Lordosis Angle | numeric, float64 | (°) |
| | $X_4$ | Sacral Slope | numeric, float64 | (°) |
| | $X_5$ | Pelvic Radius | numeric, float64 | (mm) |
| | $X_6$ | Degree | numeric, float64 | (°) |

| | | Spondylolisthesis | | |
|---|---|---|---|---|
| | $X_7$ | Pelvic Slope | numeric, float64 | (°) |
| | $X_8$ | Direct Tilt | numeric, float64 | (°) |
| | $X_9$ | Thoracic Slope | numeric, float64 | (°) |
| | $X_{10}$ | Cervical Tilt | numeric, float64 | (°) |
| | $X_{11}$ | Sacrum Angle | numeric, float64 | (°) |
| | $X_{12}$ | Scoliosis Slope | numeric, float64 | (°) |
| *Class_att (Y)* | | Attribute Class | categorical, object | 0=Abnormal, 1=Normal |

Apply independent sample *T* test descriptive statistics to find out what significantly contribute to the outcome of NSLBP. Binary logistic regression analysis was used to predict the relationship between the dependent variable (Y) and the independent variable (X). In this study, Y stands for Class_att (Abnormal or Normal). And 12 numeric predictors of X are list in *Tab.1*. To build a more accurate prediction model, Multilayer Perceptron (MLP) is applied. The MLPs breaks this restriction and classifies datasets by using a more robust and complex architecture to learn regression and classification models for difficult datasets. The MLP procedure produces a predictive model for NSLBP based on the values of the predictor variables in the regression equation.

**Spinopelvic parameters Measurement**

Optimal position for radiologic measurement of lordosis is standing with arms supported while shoulders are flexed at a 30° angle[19]. The digitized thoracic points on the lateral radiographs were all vertebral body corners of T1–T12. Subjects held onto a vertical pole with hands at elbow level to keep the upper extremities from projecting over the spine.

First interpret the lumbar X-rays and determine the degree of lumbar lordosis. Then determine the lumbar curve's Cobb angle from an X-ray taken in profile, using the centroid, tangential radiologic assessment of lumbar lordosis method (TRALL),or using the Harrison posterior tangent line-drawing methods(*Fig.1*)[20-22].

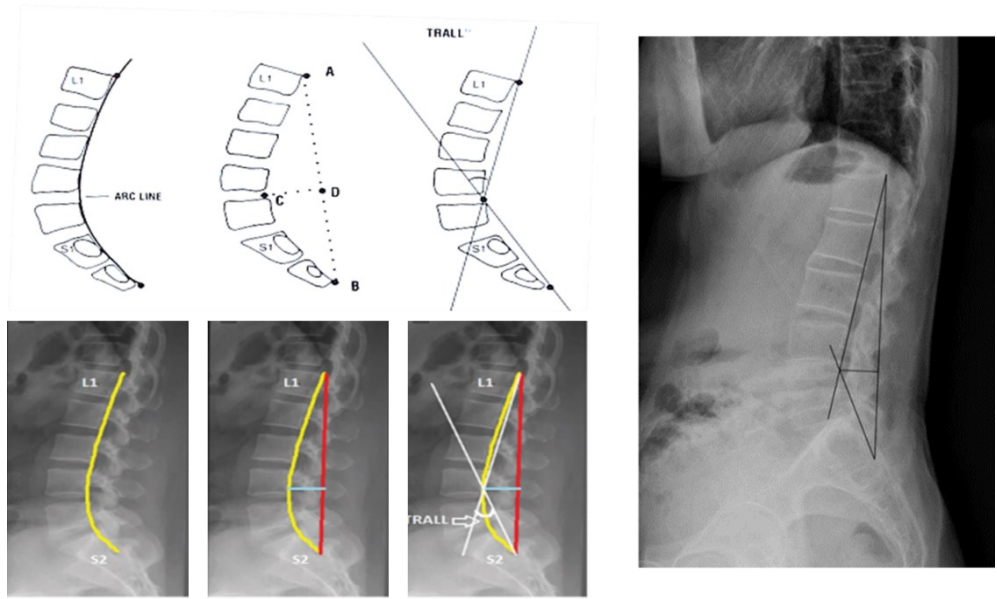

Fig. 1 *tangential radiologic assessment of lumbar lordosis method (TRALL)*

Centroid method, Cobb method and the posterior tangent method are three different radiographic analysis methods(*Fig.2*)[23]. 1)The Centroid method use four digitized body corners to construct the intersection (centroid) of vertebral body diagonals. The centroid method requires three adjacent vertebrae to construct segmental angles and either three or four vertebrae to construct global angles(*Fig.2-A*). 2) Cobb method use the inferior vertebral body corners on each thoracic segment were used to construct segmental Cobb angles (e.g., CobbT1–T2). Segmental and global Cobb angles are constructed with lines drawn on vertebral body endplates. The posterior tangent method uses the superior- posterior and inferior-posterior body corners(*Fig.2-B*)[21].3)The posterior tangent method uses the two posterior vertebral body corners. Lines are drawn tangent to the midposterior vertebral body through these two points. These lines are the slopes in an engineering analysis of columns. Relative rotation angles (segmental angles) are created by intersecting adjacent tangents. Absolute rotation angles (global angles) are constructed by intersecting tangents on the cranial and caudal segments of the curve . Global angles are sums of the intervening segmental angles(*Fig.2-C*).

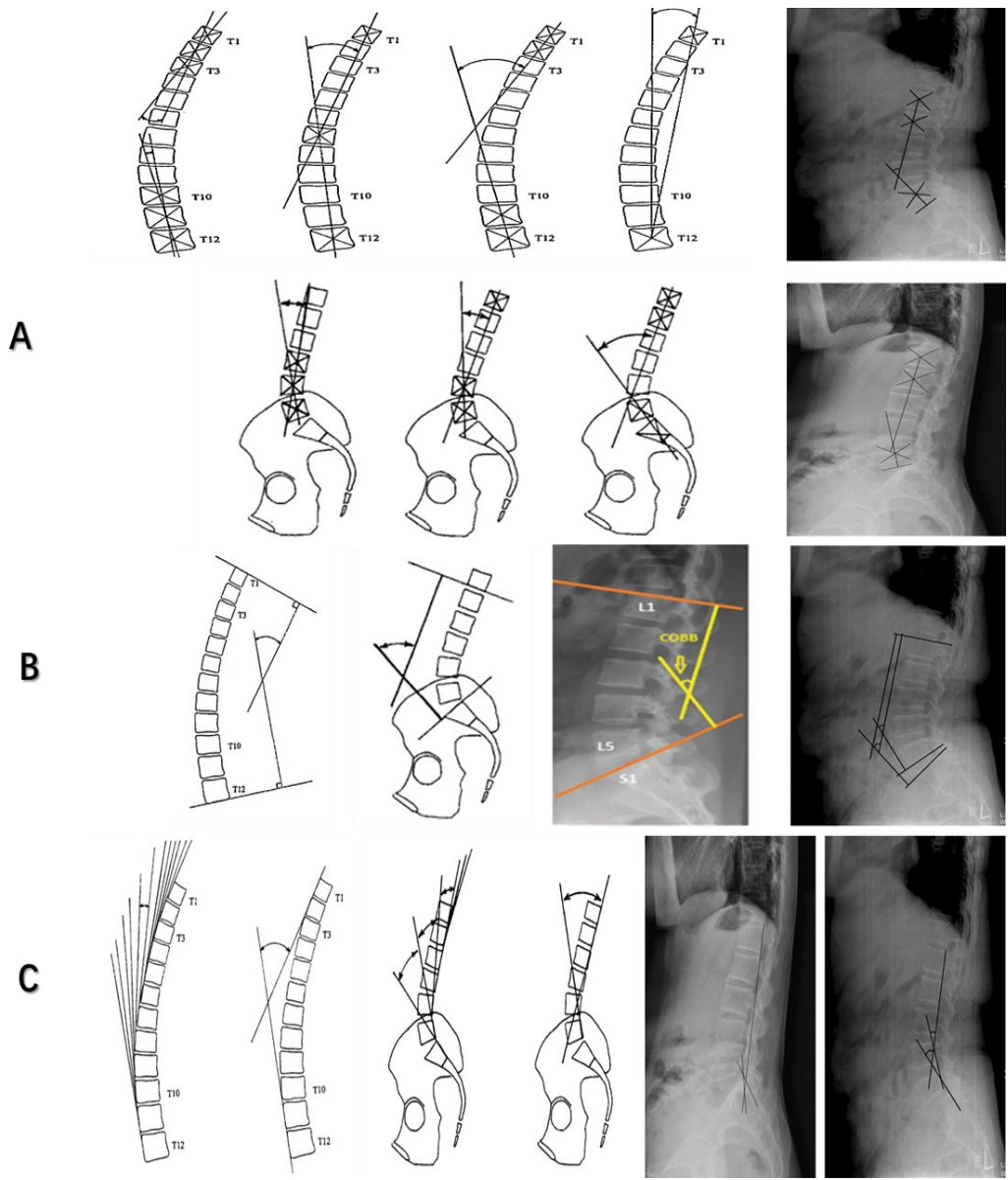

Fig. 2　*Different radiographic analysis methods*
A:*centroid method.*　B: *Cobb method.*　C: *The posterior tangent method.*

## Results

　　Spinopelvic parameters of normal are significantly different from the abnormal in Pelvic incidence (51.69±12.37 *vs*. 64.69±17.66,*F*=17.77,*P*=0.00), Pelvic tilt (12.82±6.78 *vs*.19.79±10.52, *F*=15.85,*P*=0.00), Lumbar lordosis angle (43.54±12.36 *vs*.55.93±19.67, *F*=26.93,*P*=0.00), Sacral slope (38.86±9.62 *vs*. 44.90±14.52, *F*=15.10,*P*=0.00), Pelvic radius (123.89±9.01 *vs*.115.08±14.09, *F*=10.95,*P*=0.00), Degree spondylolisthesis (2.19±6.31 *vs*.37.78±40.70, *F*=50.08,*P*=0.00).

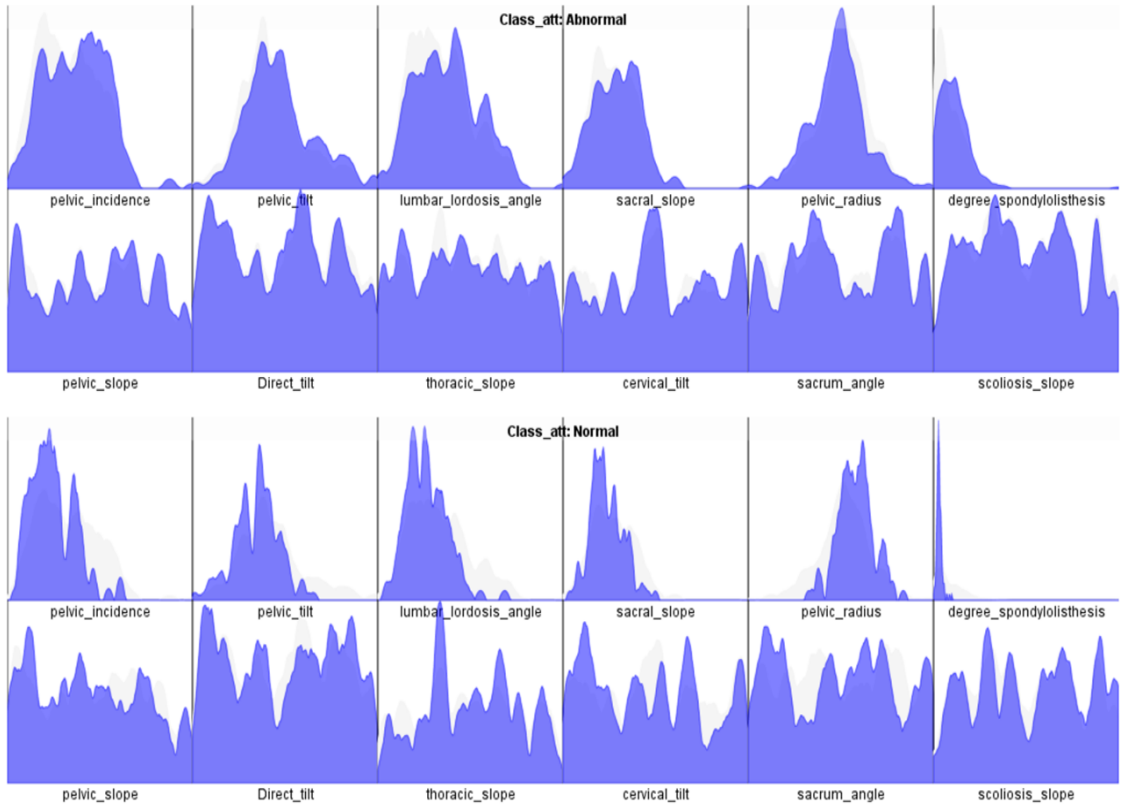

Fig. 3  *Descriptive of Spinopelvic parameters*

**Logistic regression model of NSLBP**

It can be observed that the model is statistically significant in Omnibus Tests of Model Coefficients ($\chi^2$=210.918, $P$<0.005). The Hosmer and Lemeshow goodness of fit test is not statistically significant ($P$=0.144), indicating that the model fits well. The proportion of variation that can be explained by the dependent variable is 69.0% (Nagelkerke $R^2$).

Tab. 2  *Classification table*[a]

| Observed value | | | Predictive value | | Percentage Correct |
|---|---|---|---|---|---|
| | | | class | | |
| | | | Abnormal | Normal | |
| **Step 4** | class | Abnormal | 186 | 24 | **88.6** |
| | | Normal | 20 | 80 | **80.0** |
| **Overall Percentage** | | | | | **85.8** |

[a]. The cut value is 0.500

The overall accuracy of the equation prediction model is 85.8% as shown in *Tab.2*. This model can correctly classify 85.8% of the research objects. The sensitivity of the model is 88.6% and the specificity is 80.0%. 90.3% of the observations that predicted with NSLBP were correct. And 76.9% of the observations that predicted without NSLBP were correct.

Tab. 3  *logistic Regression Predicting Likelihood of NSLBP based on PT,SS,PR,DS*

|  | B | S.E. | Wald | df | Sig. | Exp(B) | 95% CI for EXP(B) | |
|---|---|---|---|---|---|---|---|---|
|  |  |  |  |  |  |  | Lower | Upper |
| **Pelvic tilt** | -0.07 | 0.03 | 5.16 | 1 | 0.02 | 0.94 | 0.88 | 0.99 |
| **Sacral slope** | 0.11 | 0.02 | 25.72 | 1 | 0.00 | 1.12 | 1.07 | 1.17 |
| **Pelvic radius** | 0.11 | 0.02 | 22.85 | 1 | 0.00 | 1.12 | 1.07 | 1.17 |
| **Degree spondylolisthesis** | -0.17 | 0.02 | 52.24 | 1 | 0.00 | 0.85 | 0.81 | 0.89 |
| **Constant** | -15.46 | 3.27 | 22.30 | 1 | 0.00 | 0.00 |  |  |

The method of selecting variables in this statistical process is "Forward: LR" method. The variables in the equation table lists the variables and their parameters that are finally screened into the model(*Tab.3*). The "Sig." column represents the *P* value of the corresponding variable in the model, and "Exp (B) and 95% CI for EXP (B)" represent the *OR* value of the corresponding variable and its 95% confidence interval. Research subjects with higher values in " Sacral slope " and " Pelvic radius " both had 1.12 times risk of low back pain. And these two parameters both increased risk of low back pain is significant(*OR*=1.12, 95% *CI*: 1.07-1.17,*P*=0.00).

**Multilayer Perceptron Model of NSLBP**

Select the one dependent variable of *Class_att* and four predictors in the regression equation. Specify 30% of the sample to prevent overfitting the sample set in the setting before proceed. And randomly allocate 71.3% of the samples to the training set and 28.7% to the test set in this MLP model. The predictive value of overall percentage in training dataset is 88.7%,while is 87.6% in testing dataset. According to the ROC curve, the areas of "Abnormal" and "Normal" in the area model below the curve are both 0.952, indicating that the model has good predictive ability. The importance of influence of the parameters in the model on the occurrence of NSLBP is ranked as follows(*Tab.4*): Degree spondylolisthesis(100%), Pelvic radius(45.9%), Sacral slope(40.1%), Pelvic tilt(21.4%).

Tab. 4  *Variable importance*

| | Importance | Normalized importance |
|---|---|---|
| **Pelvic tilt** | 0.103 | 21.4% |
| **Pelvic radius** | 0.221 | 45.9% |
| **Degree of spondylolisthesis** | 0.482 | 100.0% |
| **Sacral slope** | 0.193 | 40.1% |

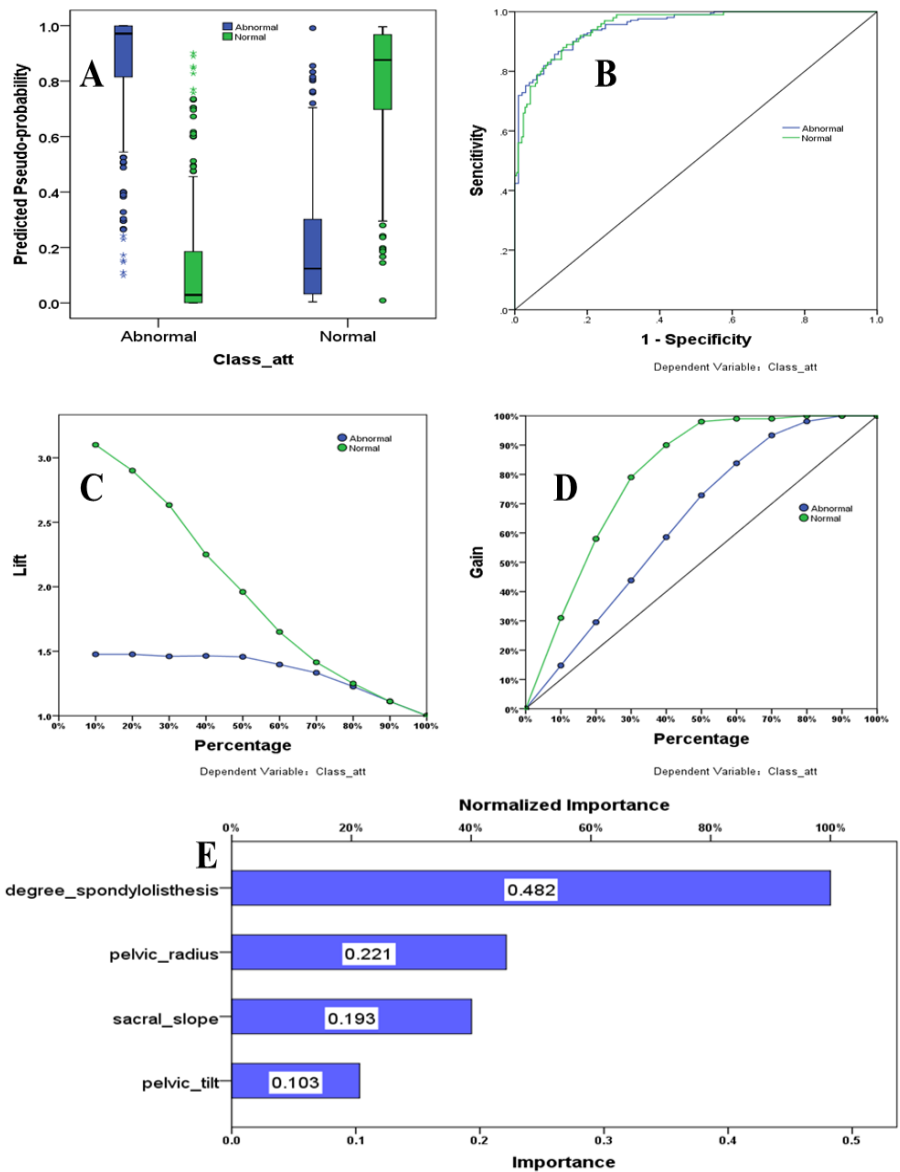

Fig. 4  *MLP prediction model for NSLBP*
*A: graph of predicted pseudo-probability; B: ROC curve; C: Gain graph;*
*D: Lift graph; E: Graph of Independent Variables Importance*

## Discussion

**Predictive model evaluation**

  In this study, the regression model screened out four predictors that can affect the occurrence of NSLBP, and the model accuracy rate was 85.8%. In the regression model, the four predictors can promote or lower the risk of NSLBP, and the positive and negative effects are different. While in the MLP model, the order of the influence of each predictor affects the accuracy of the prediction model. In the MLP prediction model established by the predictor, spondylolisthesis is the most predictive factor that determines the occurrence of NSLBP, and the model accuracy rate reaches 95.2%, suggesting The MLP model is more accurate in predicting multiple factors.

**Spinopelvic parameters and NSLBP**

Spondylolisthesis refers to translation of 1 vertebral segment compared with the subadjacent level, which can be described according to its degree of severity, causing mechanical or radicular symptoms or pain. Meyerding classification is accurate for measuring slip percentage, graded according to degree of slippage; based on the ratio of the overhanging part of the superior vertical body to the anterio-posterior length of the inferior vertebral body[24]. It is found that there is a huge difference between normal and abnormal lumbar spondylolisthesis. It shows that it plays an important and decisive role in predicting the NSLBP model. This fact tells us that most of the causes of NSLBP may come from lumbar spondylolisthesis, and vice versa.

Pelvic radius (PR) the distance from the hip axis (located in the middle between the two femoral bead mid-points[25]) to the posterior-superior corner of the S1 endplate, which the standard values was $137 \pm 9$ mm[26]. The pelvic radius parameter is also significantly different in the normal(123.89±9.01)and abnormal(115.08±14.09) category models in our study. In the regression model, for every 1mm increase in PR, the incidence of NSLBP increases by 1.12 times. In the MLP prediction model, the weight for predicting the occurrence of NSLBP accounts for 45.9%. PR is related to the stability of the pelvic space structure and also determines the balance of the spine. The sagittal balance of the spinopelvic is defined by the parameters based on notable biomechanical forces involved in the transmission of constraints with the broadening and verticalization of the pelvis and the upright position characteristic of the spinal curves structured, and the supporting muscles modified.

Sacral slope (SS) is an angle subtended by a line parallel to the sacral end plate and a horizontal reference line[27]. The normal range of value for the SS was from -32° to -49°[28].The angle between the superior plate of S1 and the horizontal reference line with a normal range from 36-42°. Normal(38.86±9.62) *vs.* abnormal (44.90±14.52)of SS in this study is significantly different according to the analyzing. And the range of normal SS in our study meets the standard of the ideal spinopelvic parameter for eliminating residual pain and disability in adult spinal deformity, which is around 30 degree[29]. The anatomical orientation of the pelvis with a high SS was one of the predisposing factors for degenerative spondylolisthesis which leads to NSLBP[30].

The pelvic tilt (PT) is an angle measured by a vertical reference line from the center of the femoral head and a line from the center of the femoral head to the midpoint of the sacral end plate[27]. The (anterior or posterior) pelvic tilt describes here the angle between the anterior pelvic plane and the coronal plane of the body[13]. Significant differences in pelvic tilt were found in this study between people with and without NSLBP (12.82±6.78 *vs.*19.79±10.52), which indicates that the evaluation of radiographic spinopelvic parameter is more accurate comparing to the measurement of individual related motion and posture captured by wearable sensors[31]. In addition, our analyzed outcomes are consistent with those of patients treated with minimally invasive surgical treatment of transforaminal lumbar interbody fusion[32], that is, a greater decrease in PT is associated with an improvement in back pain.

## Conclusion

DS,PR,SS,PT are four predictors screened out by regression analysis that have significant predictive power for the risk of NSLBP. The multi-layer perceptron network algorithm determines that DS is the most powerful predictor of NSLBP through precise modeling. Through data analysis and modeling, accurate screening of pelvic spine parameters that affect NSLBP can help prevent and treat patients with NSLBP more quickly.

This method uses the NSLBP open database for analysis, and further application in clinical should be the next step ahead. Although computer science has its strong advantages in data analysis, its application in the field of medical clinical research requires more verification and screening.

## Acknowledgements

Thanks are due to Science and Technology Bureau (Zhanjiang City) technological guidance special project: A Randomized Controlled Trial Study of Physical Activity and Level of Participation Changes in a Population with Lower Back Pain Under an Exercise Guidance Intervention，No. 2017A01014.